\documentclass[aps,pre,twocolumn, titlepage]{revtex4}
\usepackage{graphicx}
\usepackage{color}
\usepackage{dcolumn}
\usepackage{amsmath}
\usepackage{amsfonts}
\usepackage{bm}
\usepackage{epstopdf}

\begin{document}
\title{Dynamical commensuration effect in a two-dimensional Yukawa solid modulated by periodic substrates}
\author{Wenqi Zhu$^1$, C. Reichhardt$^2$, C. J. O. Reichhardt$^2$, and Yan Feng$^1,$ $^{3,}$ $^\ast$}
\affiliation{
$^1$  Institute of Plasma Physics and Technology, School of Physical Science and Technology, Soochow University, Suzhou 215006, China\\
$^2$ Theoretical Division, Los Alamos National Laboratory, Los Alamos, New Mexico 87545, USA\\
$^3$ National Laboratory of Solid State Microstructures, Nanjing University, Nanjing 210093, China\\
$\ast$ E-mail: fengyan@suda.edu.cn}
\date{\today}

\begin{abstract}
Transverse depinning dynamics of a periodic-square-substrate modulated two-dimensional dusty plasma solid driven by a constant force in the longitudinal direction are investigated using Langevin dynamical simulations. During the increase of the commensuration ratio (the number ratio of particles to substrate potential wells), the nonmonotonic variation trend of the critical transverse depinning force is observed. It is found that the local maxima and minima of the critical transverse depinning force just correspond to the dynamical commensurate and incommensurate conditions, respectively. The dynamical commensurate effect is also clearly exhibited from the stable one-dimensional channel particle trajectories and the highly ordered structure, however, both the particle trajectories and the structure are more disordered under the incommensurate conditions. The nonmonotonic variation of the critical transverse depinning force is attributed to the stability of the lattice structure under various commensuration ratios. 

\end{abstract}

\maketitle

\section{Introduction}

While interacting particles modulated by substrates are driven by an external force, depinning dynamics occur~\cite{Reichhardt:2017,Bhattacharya:1993,Tierno:2012,Williams:1991}, which have been widely investigated in numerous two-dimensional (2D) physical systems~\cite{Reichhardt:2017}, such as vortices in type-II superconductors~\cite{Bhattacharya:1993,Koshelev;1994,Pardo:1998,Olson:1998}, colloids~\cite{Pertsinidis:2008,Bohlein:2012,Tierno:2012,Tang:1994,Ertas:1994}, Wigner crystals~\cite{Williams:1991,Cha:1998}, and pattern-forming systems~\cite{Reichhardt:2003,Sengupta:2010}. Depinning dynamics contain the fundamental physics significance, where various phase transitions with different properties are characterized by order parameters~\cite{Reichhardt:2017,Gu:2020}. In depinning dynamics, depinning transition occurs at a critical external driving threshold under an underlying substrate~\cite{Reichhardt:2017,Li:2019,Gu:2020}. When the external driving force is above this threshold, some or all of the particles move across the substrate, leading to abundant dynamical behaviors. The depinning dynamics may be either elastic~\cite{Scala:2012} or plastic~\cite{Fily:2010}, depending on the topological order of particles. During the depinning transition procedure, three dynamical states are typically observed, which are the pinned state, the disordered plastic flow state, and the moving ordered elastic state~\cite{Reichhardt:2017,Li:2019,Gu:2020}. While driving various physical systems modulated by assorted 2D substrates using different magnitudes of the external force, abundant dynamical responses have been observed in the depinning dynamics, such as the superlubricity transition~\cite{Huang:2022,Mandelli:2015}, the directional locking~\cite{Xiao:2010,Tierno:2007,Lacasta:2005,Reichhardt:1999,Reichhardt:2004,Reichhardt:2011,Reichhardt:2012,Reichhardt:20122,Zhu:2022}, the continuous/discontinuous phase transitions~\cite{Reichhardt:2017,Gu:2020}, and the dynamical commensuration effect~\cite{Reichhardt:2011,Reichhardt:2008,Reichhardt:2012,Reichhardt:20122}, as studied here. The commensuration ratio refers to the ratio of the total particle number to the number of the substrate potential well~\cite{Reichhardt:1999,Reichhardt:2011}.

For interacting particles modulated by 2D susbtrates and driven to move along the longitudinal direction at first, when the applied transverse driving force gradually increases from zero, the critical transverse depinning force, which starts triggering the transverse motion, exhibts a nonmonotonic variation trend with the commensuration ratio~\cite{Anders:2000,Karapetrov:2005}, which is associated with the dynamical commensuration effect~\cite{Reichhardt:2011,Reichhardt:2008,Reichhardt:2012,Reichhardt:20122}. In~\cite{Reichhardt:2011,Reichhardt:2008}, the uniform stable one-dimensional (1D) channel trajectories and the highly ordered structure are observed in the dynamical commensuration states. The dynamical commensuration effect is typically studied with the directional locking effect in overdamped systems~\cite{Reichhardt:1999,Reichhardt:2011,Reichhardt:2012,Reichhardt:20122}, since both of them occur while varying the external driving force and the 2D substrate modulation. The directional locking effect is reflected by the locking steps~\cite{Reichhardt:1999,Reichhardt:2012}, corresponding to the ``locked'' collective velocity direction, which occur when the direction of the external driving force is close to the symmetry direction of the substrate. While the dynamical commensuration effect is exhibited by the dynamical commensurate and incommensurate conditions which occur at different commensuration ratios~\cite{Reichhardt:2011,Reichhardt:2008,Reichhardt:20122}. Since the directional locking effect has just been studied in a periodic-square-substrate modulated 2D dusty plasma recently~\cite{Zhu:2022}, here we mainly focus on the dynamical commensuration effect in this system.

As an excellent model system, dusty plasma, or complex plasma, refers to the four component mixture of micron-size dust particles, free electrons, free ions, and neutral gas atoms~\cite{Thomas:1996,I:1996,Chu:1994,Morfill:2009,Thomas:1994,Fortov:2005,Piel:2010,Bonitz:2010,Melzer:1996,Merlino:2004,Feng:2008,Thomas:2004}. In typical experiments, by absorbing free electrons and ions, these dust particles are charged to a high negative charge of $\sim -{10}^{4} e$, leading to the strong coupling between these particles, as a result, they exhibit typical collective behavior of solids~\cite{Feng:2008,Hartmann:2014} and liquids~\cite{Thomas:2004,Feng:2010}. In experiments, a single layer suspension of dust particles can be formed, also called 2D dusty plasma~\cite{Feng:2011,Qiao:2014}, where the interaction between particles can be modeled as the Yukawa repulsion~\cite{Konopka:2000}. Due to the low gas pressure of the plasma~\cite{Feng:20102,Feng:2012,Kananovich:2020}, the particle motion is underdamped~\cite{Liu:2003}. Since these dust particles can be directly identified using video imaging and their motion can be also tracked in captured movies~\cite{Feng:20163}, various fundamental physics procedure can be studied at the individual particle level using dusty plasmas.

Recently, collective behaviors of 2D dusty plasmas modulated by various substrates have been extensively investigated using computer simulations~\cite{Li:2018,Huang:2022,Wang:2018,Feng:2021,Li:2019,Gu:2020,Zhu:2022}. While driving the substrate modulated 2D dusty plamsa using an increasing external driving force, three dynamical states of the pinned, the disordered plastic flow, and the moving ordered elastic states are observed~\cite{Li:2019,Gu:2020}. The phase transitions between these states are either continuous or discontinuous, which depends on the depth of the substrate wells~\cite{Gu:2020}. When the lattice structure and the underlying 2D substrate are mismatched, particles are able to move freely on a substrate, eventhough the driving force is extremely tiny, i.e., the typical supperlubricity are observed~\cite{Huang:2022}. When the direction of the external driving force is close to the symmetry direction of the substrate, the direction of the particle motion is locked in this symmetry direction, i.e., the directional locking effect is found~\cite{Zhu:2022}. However, from our literature, the dynamical commensuration effect has not been studied in dusty plasmas yet, as studied here. 

In most systems where dynamical locking effects and depinning in the presence of commensuration effects have been considered, overdamped dynamics were typically used. A new aspect to studying these effects with dusty plasmas is that the dynamics is underdamped, so that the inertial effects come into play. Our work shows that dynamical commensuration effects are robust for systems with the inertial effects, which could be relevant to other systems interacting with a periodic substrate, such as levitated colloids, cold atoms, and atomic friction, where the inertia is relevant.

In this paper, the rest portions are organized as follows. In Sec.~\uppercase\expandafter{\romannumeral2}, we briefly introduce our simulation methods to mimic periodic-square-substrate modulated 2D dusty plasma driven by an external force. In Sec.~\uppercase\expandafter{\romannumeral3}, we report the nonmonotonic variation of the critical transverse depinning force with the commensuration ratio. We also present our observed dynamical commensuration effect associated with the stable 1D particle trajectories and the highly ordered structure. In Sec.~IV, we provide a summary of our findings.

\section{Simulation methods}

To investigate the dynamical commensuration effect of a periodic-square-substrate modulated 2D dusty plasma driven by an external force, we perform Langevin dynamical simulations~\cite{Li:2018}. The interaction between particles is assumed to be the Yukawa repulsion $\phi_{ij} = Q^2 {\rm exp}(-r_{ij} / \lambda_D) / 4 \pi \epsilon_0 r_{ij}$~\cite{Konopka:2000}, where $Q$ is the charge on each particle, $r_{ij}$ is the distance between the $i$th and $j$th particles, and $\lambda_ D$ is the Debye length. Our simulated dusty plasma system is traditionally characterized by the coupling paremeter $\Gamma$ and the screening parameter $\kappa$~\cite{Fortov:2005,Morfill:2009,Piel:2010,Bonitz:2010}, defined~\cite{Ohta:2000,Sanbonmatsu:2001} as $\Gamma = Q^2/(4 \pi \epsilon_0 a k_{B} T)$ and $\kappa = a / \lambda_{D}$, respectively, where $a = 1/\sqrt{\pi n}$~\cite{Kalman:2004} is the Wigner-Seitz radius for the areal number density $n$ and $T$ is the averaged kinetic temperature of particles.

In our Langevin dynamical simulations, the equation of motion for the $i$th particle is
\begin{equation}\label{Langevin}
{	m \ddot{\bf r}_i = -\nabla \Sigma_j \phi_{ij} - \nu m\dot{\bf r}_i + \xi_i(t)+{\bf F}_s+{\bf F}_d }.
\end{equation}
Here, the five terms on the right-hand-side just come from the interparticle Yukawa repulsion, the frictional gas damping~\cite{Liu:2003}, the Langevin random kicks~\cite{Feng:2008,Feng:20082}, the force from the substrate, and the external driving force, respectively. In total, there are $N = 4096$ simulated particles constrained in a rectangular 2D box of $121.9a\times105.6a$ with the periodic boundary conditions. 

In our simulations, the applied square substrate is assumed to be~\cite{Lacasta:2005}
\begin{equation}\label{substrate}
U(x, y)=\frac{-V_{0}}{1+e^{-g(x, y)}},
\end{equation}
leading to the force of ${\bf F}_{s}$, the same as Eq.~(3) of~\cite{Zhu:2022}, where $V_0$ controls the depth of potential wells of the square substrate in units of $E_0=Q^2/4\pi\epsilon_0 a$. Here, $g(x, y)=A\left[\cos \left(2 \pi x / w_{x}\right)+\cos \left(2 \pi y / w_{y}\right)-2 B\right]$~\cite{Lacasta:2005}, where $w_x$ and $w_y$ are the spatial periods of the wells in the $x$ and $y$ directions, while $A$ and $B$ determine the steepness and the relative size of the potential wells, respectively. In our current study, $V_0$, $A$, and $B$ are specified as constant values of $V_0=0.1E_0$, $A=6$, and $B=0.8$, respectively. However, the values of $w_x$ and $w_y$ are varied simultaneously to adjust the total number of potential wells of the square substrate $N_w$, leading to the varying commensuration ratio $\rho = N/N_w$ from $0.503$ to $6.095$ for our simulated 2D Yukawa solid. Note that, to satisfy the assumed periodic boundary conditions in our 2D simulation box, $w_x$ and $w_y$ are not exactly the same, containing $< 1.5\%$ difference. For the external driving force ${\bf F}_d=F_x\bm{\hat{x}}+F_y \bm{\hat{y}}$, in units of $F_0=Q^2/4 \pi \epsilon_0 a^2$, to study the transverse depinning dynamics during the motion in the longitudinal $x$ direction, we set the unchanged driving force in the $x$ direction as $F_x=0.24F_0$, while increase $F_y$ gradually from $0$ to $0.03F_0$, for each specified commensuration ratio. 

Here are some other parameters in our simulations. The conditions of 2D Yukawa system are specified as $\Gamma=1000$ and $\kappa=2$, corresponding to the typical solid state~\cite{Hartmann:2005}. The frictional gas damping rate is chosen to be similar to the typical experimental value~\cite{Feng:2011} as $\nu = 0.027 {\omega}_{pd}$, where $\omega_{pd}={(Q^2/2 \pi \epsilon_0 m a^3)}^{1/2}$ is the nominal 2D dusty plasma frequency~\cite{Kalman:2004}. As justified in~\cite{Liu:2005}, the time step to integrate Eq.~(\ref{Langevin}) is chosen to be $0.0014\omega_{pd}^{-1}$. For each simulation run, after the steady state is achieved, we integrate Eq.~(\ref{Langevin}) for all particles at least $\ge 1.3\times{10}^6$ steps to obtain their positions and velocities for our data analysis reported next.

\section{results and discussions}

\subsection{Results}

To study the effect of the commensuration ratio $\rho$ on the transverse depinning dynamics~\cite{Reichhardt:2008}, we calculate the collective transverse drift velocity $V_y$ as the transverse driving force $F_y$ increases from zero, for various values of the commensuration ratio $\rho$, as plotted in Fig.~1. As mentioned above, at first, we fix the external force in the $x$ direction as $F_x=0.24F_0$, so that our simulated 2D Yukawa solid always slides in the $x$ direction at a constant speed. Next, we choose one value of $F_y$ from $0$ to $0.03F_0$ to perform one Langevin simulation run to reach the steady state, then continue the simulation with $\ge 1.3\times{10}^6$ steps to record the positions and velocities of all particles. By changing the $F_y$ value and repeating the similar simulation procedure, the detailed transverse depinning dynamics under various conditions are achieved. Finally, we calculate the collective velocity projected in the $y$ direction $V_{y}$ in units of $V_0=(Q^2/4\pi\epsilon_0 m a)^{1/2}$ using ${V_{y}=N^{-1}\left\langle\sum_{i=1}^{N}\dot{\bf r}_i \cdot \bm{\hat{y}}\right\rangle}$, where $\left\langle~\right\rangle$ represents the ensemble average for all particles, as plotted in Fig.~1.

From the obtained transverse drift velocity of the periodic-square-substrate modulated 2D Yukawa solid presented in Fig.~1, three typical dynamical states are clearly exhibited in the transverse depinning procedure, which are the pinned state, the disordered plastic state, and the moving ordered state~\cite{Li:2019}. For all of the four commensuration ratio values, when the transverse driving force $F_y$ is small, the magnitude of $V_y$ is always nearly zero, suggesting that particles are pinned in the $y$ direction. However, when $F_y$ is large enough, for all $\rho$ values, $V_y$ always increases linearly with $F_y$ with the constant slope of $1/\nu m$, which is a typical property of the moving ordered state~\cite{Li:2019, Feng:2021}. When $F_y$ is in the moderate range between the typical pinned and moving ordered states, corresponding to the typical plastic flow state, the variation of $V_y$ with $F_y$ is completely different for these four $\rho$ values. When $\rho = 2.032$ and $6.095$, the gradual increase of $V_y$ in Fig.~1 clearly indicates that the depinning transition is continuous. However, when $\rho = 1.089$ and $3.900$, the sudden jump of $V_y$ in Fig.~1 indicates that the discontinuous depinning transition occurs.

To further characterize the depinning transition, we use the diagnostic of the critical transverse depinning force $F_y^c$~\cite{Reichhardt:1999}. The critical transverse depinning force $F_y^c$ is defined as the largest transverse driving force, within which all particles are still pinned in the $y$ direction~\cite{Reichhardt:1999}. For the discontinuous depinning transition, the boundary between the pinned and depinning states is distinctive from the variation of $V_y$, so that the value of $F_y^c$ can easily identified in Fig.~1, which are $0.017F_0$ and $0.012F_0$ for $\rho=1.089$ and $3.900$, respectively. For the continuous depinning transition, since $V_y$ increases gradually, a threshold value of $V_y$ is needed to define the critical transverse depinning force $F_y^c$. Here, we choose the threshold value of $V_y=1\times 10^{-4}V_0$, so that the corresponding $F_y^c$ values are $0.007F_0$ and nearly $0$, for $\rho=2.032$ and $6.095$, respectively, as the two magnified insets shown in Fig.~1. 

\begin{figure}[htb]
\centering
\includegraphics{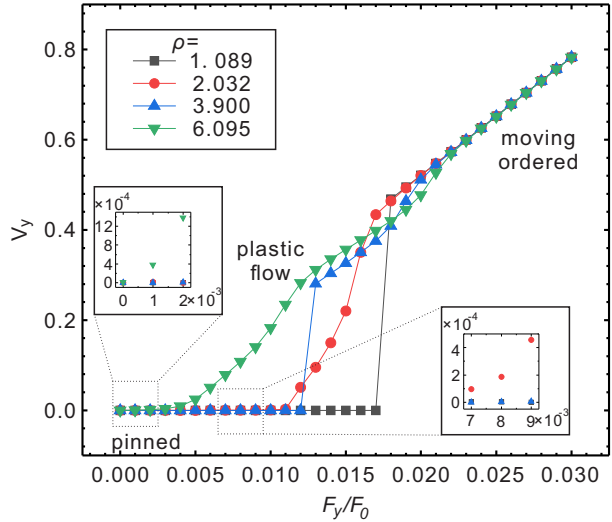}
\caption{\label{depinning} Obtained collective transverse drift velocity $V_y$ versus the transverse driving force $F_y$ for a square-substrate modulated 2D Yukawa solid, with various values of the commensuration ratio $\rho=1.089$, $\rho=2.032$, $\rho=3.900$, and $\rho=6.095$, respectively. Here the driving force in the $x$ direction is unchanged at $F_x=0.24F_0$, where $F_0=Q^2/4 \pi \epsilon_0 a^2$. As the transverse driving force $F_y$ gradually increases from zero, at first the dust particles are pinned in the $y$ direction until the transverse force $F_y$ reaches the critical transverse depinning force $F_y^c$. For the conditions of $\rho=1.089$, $\rho=2.032$, $\rho=3.900$, and $\rho=6.095$, the obtained values of the critical transverse depinning force $F_y^c$ from our investigations are $0.017~F_0$, $0.007~F_0$, $0.012~F_0$, and nearly zero, respectively. Clearly, our obtained values of $F_y^c$ for the conditions of $\rho=1.089$ and $\rho=3.900$ are substantially higher and more significant than those for $\rho=2.032$ and $\rho=6.095$.
}
\end{figure}

To further investigate the detailed transverse depinning dynamics under all of the conditions in our simulations, we present the determined critical transverse depinning force $F_y^c$ for the varying commensuration ratio values in Fig.~2(a). Clearly, as the commensuration ratio $\rho$ increases from 0.503 to 6.095, the obtained critical transverse depinning force $F_y^c$ varies nonmonotonically, exhibiting a few maxima and minima. Our obtained $F_y^c$ curve in Fig.~2(a) exhibits two well-defined local maxima, one is sharper centered around $\rho=1.089$, while the other is broader centered around $\rho=3.900$. Furthermore, in Fig.~2(a), two prominent local minima are also exhibited around $\rho=2.032$ and $6.095$, respectively. Note, our choice of the threshold value of $V_y=1\times 10^{-4}V_0$ has an effect on the variation of $F_y^c$ as the function of $\rho$, i.e., a higher threshold leads to reduced ripples in $F_y^c$, while a lower threshold results in a magnified oscillation in $F_y^c$, although the general trend is similar.

To investigate the mechanism of the $F_y^c$ variation with the commensuration ratio $\rho$, we calculate the friction of sixfold coordinated particles $P_6$ of our studied 2D Yukawa solid when $F_y = 0$~\cite{Reichhardt:2011}, as presented in Fig.~2(b). Here, $P_6$ is calculated by $\textstyle{P_{6}=N^{-1}\left\langle\sum_{i=1}^{N} \delta\left(6-z_{i}\right)\right\rangle}$, where the coordination number of the $i$th dust particle $z_i$ is obtained from the corresponding Voronoi construction~\cite{Li:2019}. From Fig.~2(b), as the commensuration ratio $\rho$ increases, the calculated $P_6$ value increases and decreases back and forth. Clearly, at the initial data point at $\rho=0.503$, the calculated $P_6$ is only about $0.539$, indicating a highly disordered state. Next, as $\rho$ increases gradually, $P_6$ increases to the local maximum of about $0.95$ centered at $\rho=1.089$, which means the system is in a highly ordered triangular lattice. Then, as $\rho$ further increases, $P_6$ decreases to the local minimum of $\textless0.6$ around $\rho=2.032$, corresponding to a disordered state. As the $\rho$ value increases to 6.095, our calculated $P_6$ values exhibit a few small ripples, also corresponding to ordered and disordered structures.

By comparing the results of $F_y^c$ and $P_6$ in Figs.~2(a) and 2(b) as $\rho$ increases, we find that the variation trend of $P_6$ is roughly synchronized with that of $F_y^c$~\cite{Reichhardt:2011}. Here, we label four typical conditions of the commensuration ratio $\rho$ values, corresponding conditions in Fig.~1 and also Fig.~3 presented later. Clearly, when the commensuration ratio $\rho$ increases from $1.089$ to $2.032$, both $F_y^c$ and $P_6$ decrease in general from the local maximum to the local minimum. Next, both $F_y^c$ and $P_6$ increase to the local maximum when $\rho$ increases from $2.032$ to $3.900$. Finally, as $\rho$ increases from $3.900$ to $6.095$, $F_y^c$ decays from its maximum to its minimum, and $P_6$ roughly exhibits the similar trend, although the $P_6$ feature at higher $\rho$ values is slightly different, as we will explain later.

From our understanding, the general synchronized variation trends of $F_y^c$ and $P_6$ are reasonable. In Fig.~2(b), when the obtained $P_6$ value reaches the local maximum at the specific commensuration ratio, the 2D Yukawa solid exhibits a more ordered structure, corresponding to a more stable lattice. Clearly, for a more stable lattice, a larger transverse driving force $F_y$ is needed to reach the depinning transition. As a result, the corresponding critical transverse depinning force $F_y^c$ is also higher. However, for a lower $P_6$ value, corresponding to a more unstable lattice, the resulting $F_y^c$ is reasonably lower.

Note that, when $\rho <1$, the obtained results of $P_6$ and $F_y^c$ exhibit different features, i.e., $P_6$ increases substantially, while $F_y^c$ just fluctuates briefly. The substantial increase of $P_6$ comes from the lower $\rho$ value than unity, which means there are more potential wells than particles. As a result, the calculated structural measure $P_6$ from the distribution of particles settled in more potential wells does not reflect the stability of the lattice at all. When $\rho <1$, the stability of the lattice is mainly provided by the square substrate, however, when $\rho >1$, the calculated $P_6$ does reflect the stability of the lattice itself.

\begin{figure}[htb]
\centering
\includegraphics{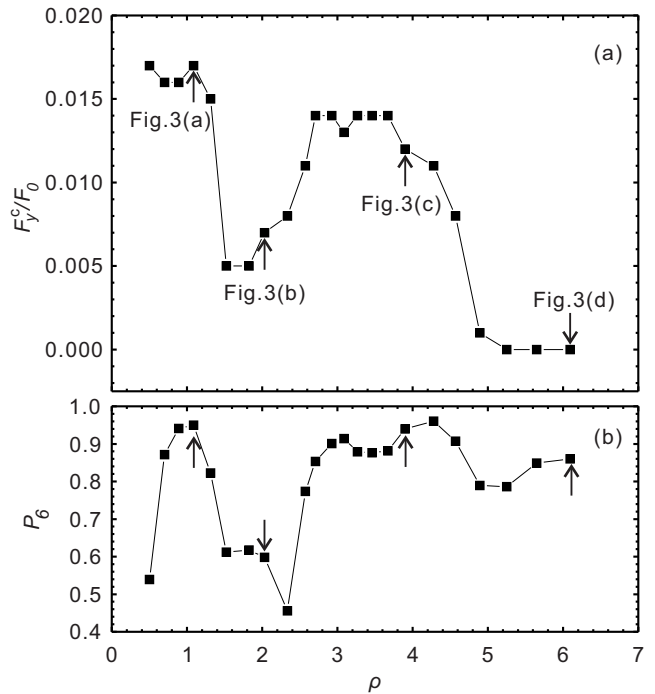}
\caption{\label{critical} Obtained values of the critical transverse depinning force $F_y^c$ from our simulations (a) and the corresponding structure measure (the fraction of sixfold coordinated particles $P_6$) at $F_y=0$ (b), for varying values of the commensuration ratio $\rho$. As the commensuration ratio $\rho$ increases, instead of increasing monotonically, the critical transverse depinnig force $F_y^c$ in panel (a) increases and decreases back and forth, exhibiting a few maxima and minima. The critical transverse depinning force $F_y^c$ reaches the local maxima around the commensuration ratio of $\rho=1.089$ and $3.900$, respectively. When the commensuration ratio $\rho=2.032$ and $6.095$, the critical transverse depinning force $F_y^c$ reaches the local minima. By comparing panels (a) and (b), it is clear that the maxima of $F_y^c$ around $\rho=1.089$ and $3.900$ just correspond to the peak values of $P_6$, while the minimum of of $F_y^c$ around $\rho=2.032$ corresponds to the dip in $P_6$. However, for the condition of $\rho=6.095$, the $P_6$ value is still high, not synchronized with the drop of $F_y^c$ any more, as explained next. 
}
\end{figure}

To explore the dynamical properties of our simulated system under these $\rho$ values, we plot the typical particle trajectories when $F_y = 0$ in Fig.~3. The four $\rho$ values in Fig.~3 are $1.089$, $2.032$, $3.900$, and $6.095$, respectively, which are the same as those in Fig.~1, also marked in Figs.~2(a, b), roughly corresponding to the local maxima and minima of $F_y^c$. Here, the plotted particle trajectories correspond to the temporal duration of $t{\omega}_{pd} = 5.6$ and the spatial region of $0.8\%$ of our total simulation box. In each panel of Fig.~3, the open circles represent the locations and the relative size of the potential wells, while the filled dots indicate the positions of the simulated particles in one typical snap shot~\cite{Zhu:2022}.

From Fig.~3, under the different values of $\rho$, the particle trajectories exhibit completely different properties, corresponding to either the dynamical commensurate or dynamical incommensurate states. In Fig.~3(a), under the condition of $\rho=1.089$, corresponding to the first local maximum of $F_y^c$, the particle trajectories are aligned in the 1D channels, always crossing potential wells of the square substrate, which are pretty uniform and stable~\cite{Reichhardt:2008,Reichhardt:2012}. When $\rho=3.900$ in Fig.~3(c), corresponding to the second broad maximum in Fig.~2(a), the particle trajectories are also aligned in the uniform and stable 1D channels, which are either along potential wells or just along the straight line located at the center between two adjacent wells in the $y$ direction~\cite{Reichhardt:2008,Reichhardt:2012}. That is to say, under the specific conditions corresponding to the local maxima of $F_y^c$, the system exhibits the typical dynamical commensuration effect, associated with the uniform and stable 1D channel trajectories, resembling the dynamical commensuration feature observed in other systems~\cite{Reichhardt:2011,Reichhardt:2008,Reichhardt:2012,Reichhardt:20122}.

Under the conditions of the commensuration ratio $\rho$ corresponding to the local minima in Fig.~2(a), it seems that the particles are not able to form commensurate 1D channel trajectories any more. When $\rho=2.032$ and 6.095, corresponding to the first and second local minima of $F_y^c$, the corresponding particle trajectories in Figs.~3(b) and 3(d) exhibit much disordered features, occupying the whole simulation box, not the uniform 1D channels any more. These trajectories just correspond to the dynamical incommensurate state, well consistent with the previous investigations in other physical systems~\cite{Reichhardt:1999,Reichhardt:2012}.

\begin{figure}[htb]
\centering
\includegraphics{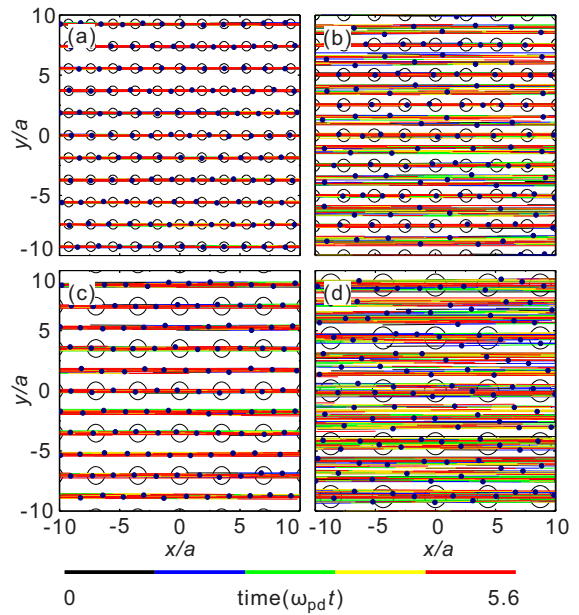}
\caption{\label{trajectories} Obtained particle trajectories of the periodic-square-substrate modulated 2D Yukawa solid for the time duration of $t \omega_{pd} = 5.6$ with the unchanged driving force of $F_x=0.24F_0$ and $F_y=0$, under the conditions of $\rho=1.089$ (a), 2.032 (b), 3.900 (c), and 6.095 (d), as marked in Fig.~2, respectively. The large open circles indicate the locations of the potential wells of the square substrates, while the small filled dots present the particle positions at one typical snap shot. From these trajectories, particles move along stable 1D channels under the conditions of $\rho=1.089$ (a) and $3.900$ (c), suggesting that the system exhibits the dynamical commensuration effect. However, when $\rho=2.032$ (b) and $6.095$ (d), the trajectories are much more disordered, where uniform and stable 1D channels disappear, so that the dynamical commensuration effect is not observed any more. 
}
\end{figure}

To further characterize the structure difference between the dynamical commensurate and incommensurate states, we calculate the 2D distribution function $G_{xy}$ of our simulated 2D Yukawa solid in Fig.~4, under the same conditions as the four panels of Fig.~3. As a powerful structural diagnostic widely used in anisotropic systems, the 2D distribution function $G_{xy}$~\cite{Loudiyi:1992} provides the possibility of finding another particle at one 2D location relative to the central particle. 

Our calculated 2D distribution functions $G_{xy}$ under various conditions exhibit significantly different features for the commensurate and incommensurate states, as shown in Fig.~4. Under the commensurate conditions of $\rho=1.089$ and $3.900$ in Figs.~4(a) and (c), the calculated 2D distribution functions clearly indicate that the structure is in the highly ordered triangular lattice, with the main axis in the $x$ direction. The highly ordered structure of $G_{xy}$ in the commensurate state of Figs.~4(a) and (c) also well agrees with the the uniform and stable 1D channel trajectories in Figs.~3(a) and (c). Note that, when the $y$ value is larger in Fig.~4(a), there is a slight smashed feature in $G_{xy}$, similar to those observed in~\cite{Zhu:2022}, probably corresponding to the particle motion in the $x$ direction. 

As compared with Figs.~4(a, c), our calculated 2D distribution functions $G_{xy}$ in the incommensurate conditions of Figs.~4(b, d) clearly indicate that the structure of our simulated 2D Yukawa solid is more disordered. While comparing the four panels of Fig.~4, the more disordered structure is reflected by the more uniformly distributed intensity in $G_{xy}$, which is more pronounced in Figs.~4(b, d). From Figs.~3(b, d), the incommensurate trajectories under the conditions of $\rho=2.032$ and $6.095$ are more disordered, reasonably consistent with the disordered structure reflected by our calculated $G_{xy}$ in Figs.~4(b, d). In Fig.~4(b), it seems that the calculated 2D distribution function $G_{xy}$ exhibits the square symmetry at a shorter distance. From $G_{xy}$ in Fig.~4(d), the ordered triangular symmetry can be still observed at a shorter distance, while this symmetric feature gradually fades away at longer distances.

\begin{figure}[htb]
\centering
\includegraphics{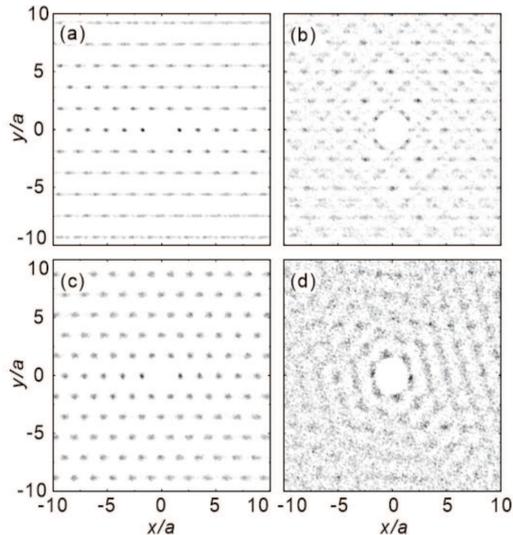}
\caption{\label{Gxy} Obtained 2D distribution functions $G_{xy}$ for the periodic-square-substrate modulated 2D Yukawa solid under the same conditions of Fig.~3. Under the commensurate conditions of $\rho=1.089$ (a) and $3.900$ (c), the calculated $G_{xy}$ indicates that the simulated system is in a highly ordered triangular lattice with the main axis along the $x$ direction, well agreeing with the dynamical commensuration effect. However, when $\rho=2.032$ (b) and $6.095$ (d), the calculated $G_{xy}$ exhibits more disorder features in the structure, agreeing with the disordered flowing rows in Figs.~3(b, d).
}
\end{figure}

\subsection{Discussions}

From our results presented above, it is clear that, under the commensurate conditions, the highly ordered structure and the higher critical transverse depinning force $F_y^c$ values occur nearly simultaneously, while under the incommensurate conditions, the more disordered structure appears with the lower $F_y^c$ value. From our interpretation, when the $P_6$ value reaches the local maxima, corresponding to the highly ordered structure of the studied 2D Yukawa solid, the stability of this highly ordered lattice reasonably leads to a higher critical transverse depinning force $F_y^c$ to destroy this lattice. In addition, under the commensurate conditions, the particle trajectories are just stable 1D channels, well agreeing with the highly ordered structure. However, when $P_6$ reaches the local minima, the structure and the particle trajectories are more disordered, so that lattice is not stable any more, reasonably leading to a much lower $F_y^c$ to reach the depinning state for the particle motion in the $y$ direction.

Although the general variation trends of $F_y^c$ and $P_6$ are nearly synchronized, this synchronized feature seems to disappear when $\rho > 5.251$. From Fig.~2, when $\rho=6.095$, i.e., roughly at the second local minimum of $F_y^c$, the $F_y^c$ value is nearly zero, while the corresponding $P_6$ value reaches its local maximum, as briefly mentioned above. We attribute this abnormal feature to the moving ordered state under the depinning threshold of $F_y=0$ when $\rho=6.095$, as shown in Fig.~1. This moving ordered state is also further confirmed by the ordered triangular lattice at a shorter distance from $G_{xy}$ in Fig.~4(d).

Here, we would like to compare the current investigation of the dynamical commensuration effect with the previous directional locking effect~\cite{Zhu:2022}, using the similar simulation method with the unchanged longitudinal driving force $F_x$. In the previous directional locking effect investigation of~\cite{Zhu:2022}, one specific commensuration ratio is specified, while the transverse driving force $F_y$ is varied gradually from zero to a value larger than $F_x$. As $F_y$ increases gradually, there are prominent locking steps when the direction of the external driving force is close to the symmetry direction of the underlying square substrate, on which the direction of the collective drift velocity of particles is ``locked'' in the symmetry direction. In our current investigation of the dynamical commensuration effect, we mainly focus on the detailed depinning dynamics under various commensuration ratio values. As the commensuration ratio $\rho$ gradually increases, we observe both the dynamical commensurate and incommensurate states, corresponding to the local maxima and local minima of the critical transverse depinning force $F_y^c$, respectively. Under the commensurate conditions, the particle trajectories exhibit stable 1D channels and the structure is highly ordered, while the particle trajectories and structure are more disordered under the incommensurate conditions.

By comparing our observed dynamical commensuration effect in our studied underdamped dusty plasma system with that of the overdamped or strongdamped physical systems~\cite{Reichhardt:2011,Reichhardt:2008,Reichhardt:2012,Reichhardt:20122}, we find that these critical transverse depinning forces exhibit the similar nonmonotonic variation with the commensuration ratio~\cite{Reichhardt:2011,Reichhardt:2008,Reichhardt:2012,Reichhardt:20122}. Furthermore, the stucture~\cite{Reichhardt:2011,Reichhardt:2012} and the dynamical property~\cite{Reichhardt:2008,Reichhardt:2012} are also quite similar for these two types of systems. For example, we also find that the dynamical commensurate conditions occur when $\rho$ is around 1 and 4 in both the underdamped and overdamped systems~\cite{Reichhardt:2012,Reichhardt:20122}, probably corresponding to the once and twice of the potential rows for the particle row number, respectively. Based on the similarity of these physical quantities, we speculate that, the inertia term in the equation of motion for the underdamped dusty plasma system probabaly does not play an important role in the observed dynamical commensuration effect. 

\section{Summary}

In summary, using Langevin dynamical simulations, we investigate the transverse depinning procedure of a periodic-square-substrate modulated 2D Yukawa solid driven by a constant longitudinal external force of $F_x$. Under different commensuration ratio values, when $F_y = 0$, the dynamical commensurate and incommensurate states are clearly observed with the particle motion only in the $x$ direction from both the particle trajectories and the static structural measures. When $\rho = 1.089$ and 3.900, the dynamical commensuration effect is clearly exhibited, associated with the uniform 1D channel particle trajectories and the highly ordered structure. While under the incommensurate states of $\rho = 2.032$ and 6.095, both the particle trajectories and the structure are more disordered. In addition, we also find that the critical transverse depinning force $F_y^c$ reaches its local maxima under the dynamical commensurate conditions, while goes to its local minima under the dynamical incommensurate conditions. We attribute this variation trend of $F_y^c$ to the stability of the lattice structure under various $\rho$ values.

\section*{ACKNOWLEDGMENTS}

The work was supported by the National Natural Science Foundation of China under Grant Nos. 12175159 and 11875199, the 1000 Youth Talents Plan, startup funds from Soochow University, and the Priority Academic Program Development of Jiangsu Higher Education Institutions, and the U. S. Department of Energy through the Los Alamos National Laboratory. Los Alamos National Laboratory is operated by Triad National Security, LLC, for the National Nuclear Security Administration of the U. S. Department of Energy (Contract No. 892333218NCA000001).

\end{document}